\documentclass[a4paper,useAMS,usenatbib]{mnras}
\usepackage{amssymb}
\newcommand{\be}{\begin{equation}}
\newcommand{\ee}{\end{equation}}
\newcommand{\ba}{\begin{eqnarray}}
\newcommand{\ea}{\end{eqnarray}}

\def\simless{\mathbin{\lower 2.5pt\hbox
   {$\rlap{\raise 4.5pt\hbox{$\char'074$}}\mathchar"7218$}}}
\def\simgrt{\mathbin{\lower 2.5pt\hbox
   {$\rlap{\raise 4.5pt\hbox{$\char'076$}}\mathchar"7218$}}}
\include{epsf}
\title[Modelling high resolution ALMA strong lenses]
{Modelling high resolution ALMA observations of strongly lensed 
highly star forming galaxies detected by 
{\it Herschel}\,\thanks{{\it Herschel} is an ESA space observatory
with science instruments provided by European-led Principal
Investigator consortia and with important participation from NASA.}}
\author[S. Dye et al.]{S. Dye,$^{1}$\thanks{E-mail: simon.dye@nottingham.ac.uk}
C. Furlanetto$^{2,1}$,
L. Dunne$^{3,4}$,
S.A. Eales$^{3}$,
M. Negrello$^{3}$,
H. Nayyeri$^{5}$,\newauthor
P.P. van der Werf$^{\,6}$,
S. Serjeant$^{7}$,
D. Farrah$^{8}$,
M.J. Micha{\l}owski$^{4,9}$,
M. Baes$^{10}$,\newauthor
L. Marchetti$^{11,12,7}$,
A. Cooray$^{5}$,
D.A. Riechers$^{13}$,
A. Amvrosiadis$^{3}$
\vspace{4mm}\\
$^{1}$School of Physics and Astronomy, Nottingham University,
University Park, Nottingham, NG7 2RD, UK\\
$^{2}$Instituto de F\'isica, Universidade Federal do Rio Grande do Sul, 
Av. Bento Gon\c{c}alves 9500, 91501-970, Porto Alegre, RS, Brazil\\
$^{3}$School of Physics and Astronomy, Cardiff University, The Parade, 
Cardiff, CF24 3AA, UK\\
$^{4}$Institute for Astronomy, University of Edinburgh, Blackford Hill, 
Edinburgh, EH9 3HJ, UK\\
$^{5}$Department of Physics and Astronomy, University of California
Irvine, Irvine, CA, USA\\
$^{6}$Leiden Observatory, Leiden University, PO Box 9513, 
NL-2300 RA Leiden, The Netherlands\\
$^{7}$School of Physical Sciences, The Open University, Walton Hall, 
Milton Keynes MK7 6AA, UK\\
$^{8}$Department of Physics, Virginia Tech, Blacksburg, VA 24061, USA\\
$^{9}$Astronomical Observatory Institute, Faculty of Physics, Adam
Mickiewicz University, ul.~S{\l}oneczna 36, 60-286 Pozna{\'n}, Poland\\
$^{10}$Sterrenkundig Observatorium, Universiteit Gent, Krijgslaan 281 S9, 
9000 Gent, Belgium\\
$^{11}$Department of Astronomy, University of Cape Town, Private Bag X3,
Rondebosch, 7701, South Africa\\
$^{12}$Department of Physics and Astronomy, University of the Western Cape,
Robert Sobukwe Road, 7535 Bellville, Cape Town, South Africa\\
$^{13}$Department of Astronomy, Cornell University, Ithaca, NY 14853, USA\\
}

\begin{document}

\date{}

\pagerange{\pageref{firstpage}--\pageref{lastpage}} 
\pubyear{2017}

\maketitle

\label{firstpage}

\begin{abstract}
  We have modelled $\sim 0.1$\,arcsec resolution ALMA imaging of six
  strong gravitationally lensed galaxies detected by the {\it
    Herschel} Space Observatory. Our modelling recovers mass
  properties of the lensing galaxies and, by determining magnification
  factors, intrinsic properties of the lensed sub-millimetre
  sources. We find that the lensed galaxies all have high ratios of
  star formation rate to dust mass, consistent with or higher than the
  mean ratio for high redshift sub-millimetre galaxies and low
  redshift ultra-luminous infra-red galaxies. Source reconstruction
  reveals that most galaxies exhibit disturbed morphologies. Both the
  cleaned image plane data and the directly observed interferometric
  visibilities have been modelled, enabling comparison of both
  approaches. In the majority of cases, the recovered lens models are
  consistent between methods, all six having mass density profiles
  that are close to isothermal.  However, one system with poor signal
  to noise shows mildly significant differences.
\end{abstract}


\section{Introduction}

The most prodigious star formation rates observed in the Universe are
located within strongly optically obscured galaxies at high redshift
\citep[e.g.,][]{alexander05,greve05,tacconi06,pope08}.  The
ultra-violet radiation emitted by their hot young stars is absorbed by
copious quantities of enshrouding dust and re-emitted in the mid- and
far-infrared (far-IR). Observations indicate that on average they are
substantially more energetic per unit mass than local star forming
galaxies and have higher star formation efficiencies
\citep[e.g.,][]{santini14}. They are also considerably more abundant
than local ultra-luminous infra-red galaxies (ULIRGs) which have
comparable bolometric luminosities \citep[e.g.,][]{chapman05,
  swinbank10,alaghband12,rowlands14}. Capturing these systems in the
midst of a high rate of assembly is of key importance for a complete
understanding of galaxy formation. Thanks to recent advances in
sub-millimetre (submm) interferometric imaging capability with
facilities such as the Atacama Large Millimetre/submillimeter Array
(ALMA), study of these high redshift submm-bright galaxies can now be
conducted with resolutions $<0.1$\,arcsec, providing vastly more
detail than was previously possible.

Strong gravitational lensing offers an additional increase in spatial
resolution, with magnification factors often in excess of 10. This
neatly complements the high lensing bias that occurs at submm
wavelengths, which makes selection of strong lens systems relatively
easy \citep{blain96,negrello07}. In this way ALMA follow-up of
significant numbers of strongly lensed far-IR sources detected in
large area surveys such as the {\it Herschel} Astrophysical Terahertz
Large Area Survey \citep[H-ATLAS;][]{eales10}, the {\it Herschel}
Extragalactic Multi-tiered Extragalactic Survey
\citep[HerMES;][]{oliver12} and the {\it Herschel} Stripe 82 Survey
\citep[HerS][]{viero14} conducted using the {\it Herschel} Space
Observatory \citep{pilbratt10} and the millimetre wavelength surveys
carried out by the South Pole Telescope \citep{carlstrom11,vieira13}
and the Planck satellite \citep{canameras15} are beginning to bring
about rapid progress in our understanding of the early stages of
galaxy formation.  In particular, the improved sensitivity of these
facilities allows study of less luminous galaxies than previously
possible, pushing down towards the main sequence of star formation
occupied by more typical star forming systems.

Not only are these surveys quickly increasing the size of current
strong lens samples \citep[e.g.,][]{wardlow13,hezaveh13,bussmann13,
calanog14,rrobinson14,bussmann15,nayyeri16,negrello17}, they are
also extending their redshift range owing to the more
favourable submm K-correction than that which occurs at shorter
wavelengths. Due to the scaling of the lensing cross-section with lens
redshift, higher redshift sources are lensed by higher redshift lenses
on average and so the extended redshift range also allows study of
lens mass profiles in galaxies at an earlier epoch, to widen the time
period over which structural evolution in lens galaxies can be
studied. Submm lens samples therefore allow the density profile slope to be
measured at earlier times when galaxies were evolving more quickly
\citep[see, for example,][]{dye14,negrello14}.

One particular measurement which has generated significant interest
owing to its simplicity and because it provides an observational
benchmark for simulations of large scale structure is that of the mass
profile of lens galaxies on scales where baryons often dominate the
mass budget \citep[i.e., on scales of the Einstein radius; see, for
example,][]{ruff11,bolton12,barnabe12,sonnenfeld15}.  The physics
governing the baryons is complex and this gives rise to significant
uncertainties in simulations. Observational characterisation of the
way in which baryons shape the central mass profile of galaxies
therefore brings valuable insight to this problem.

The more accurate lens models afforded by higher resolution submm
follow-up also bring about improvements in model-dependent source
characteristics such as luminosity, star formation rate and gas and
dust mass but also emission line ratios, source morphology and source
kinematics which are subjected to differential magnification effects
in the reconstructed source plane. A striking example of the degree to
which enhancements to our understanding of submm sources can be
made by strong lensing can be found in several studies which
recently analysed ALMA follow-up imaging of the H-ATLAS discovered
lens system SDP81 \citep[see][]{dye15,swinbank15,rybak15a,rybak15b,
  wong15,tamura15,hezaveh16,inoue16}.  These studies serve to
illustrate how high resolution submm imaging brings about a
dramatically different interpretation of the lensed source compared to
what is inferred from optical data. Whilst significant differences
between optical and submm observations, such as large offsets in flux
centroids, are not limited to lensed sources, \citep[see, for
  e.g.,][]{hodge15,chen15}, differences are expected to be more
prevalent at higher redshifts when the rate of galaxy evolution and
assembly was higher. At these redshifts, lensing efficiency and
therefore lens magnification is high, enabling much enhanced
spatial resolution for more detailed morphological study.

Techniques to reconstruct the lensed source from interferometric data
naturally divide into those which directly model the visibilities in
the uv-plane \citep[e.g.,][]{bussmann12,bussmann13,rybak15a,hezaveh16}
and those which model the cleaned data in the image plane
\citep[e.g.,][]{dye15,inoue16}.  The advantage of the latter approach
is that the reconstruction is often vastly less computationally
intensive but this comes at a price of not working with the purest
form of the data. This can in principle cause biases in the lens
modelling, especially when coverage of the uv-plane is sparse.

In this paper, we have opted to use both uv-plane and image-plane
modelling, so that comparison between both methods can be made.  We
carry out lens modelling of ALMA imaging of six galaxy-galaxy strong
lens systems originally detected by the {\it Herschel} space observatory
within H-ATLAS and the HerMES Large Mode Survey
\citep[HELMS;][]{asboth16,nayyeri16} which is an extension to the
original HerMES fields.

The layout of this paper is as follows: Section \ref{sec_data}
outlines the data. In Section \ref{sec_method} we describe the
methodology of the lens modelling.  Section \ref{sec_results} presents
the results and we summarise the findings of this work in Section
\ref{sec_summary}.  Throughout this paper, we assume the following
cosmological parameters; ${\rm H}_0=67\,{\rm km\,s}^{-1}\,{\rm
  Mpc}^{-1}$, $\Omega_m=0.32$, $\Omega_{\Lambda}=0.68$
\citep{planck13}.

\section{Data}
\label{sec_data}

\begin{table}
\centering
\small
\begin{tabular}{lcc}
\hline
ID & $z_l$ & $z_s$ \\
\hline
H-ATLAS J142413.9+022303 & 0.595$^a$ & 4.243$^b$ \\
H-ATLAS J142935.3-002836 & 0.218$^c$ & 1.026$^d$ \\
HELMS J004714.2+032454   & 0.478$^e$ & 1.190$^e$ \\
HELMS J001626.0+042613   & 0.215$^{f,g}$ & 2.509$^e$ \\
HELMS J004723.6+015751   & 0.365$^{f,g}$ & 1.441$^e$ \\
HELMS J001615.7+032435   & 0.663$^e$ & 2.765$^e$ \\
\hline
\end{tabular}
\normalsize
\caption{The six lenses systems modelled in this work with their lens
  galaxy redshifts, $z_l$, and source redshifts, $z_s$. 
  $^a$\citet{bussmann12}. $^b$\citet{cox11}. $^c$\citet{messias14}. 
  $^d$\citet{negrello17}. $^e$\citet{nayyeri16}. $^f$\citet{amvrosiadis17}.
  $^g$Marchetti et al. (in prep.).}
\label{tab_lenses}
\end{table}

The ALMA observations modelled in this paper are contained within the
ALMA dataset ADS/JAO.ALMA\#2013.1.00358.S (PI: Eales).  The ALMA
spectral setup used for each lens system is identical, comprising Band
7 continuum observations in four spectral windows, each of width
1875\,MHz centred on the frequencies 336.5, 338.5, 348.5 and
350.5\,GHz. In each spectral window, there are 128 frequency channels
giving a resolution of 15.6\,MHz. Forty two 12\,m antennas were used
with an on-source integration time of approximately 125\,s. This
results in an angular resolution of 0.12\,arcsec and an RMS of
approximately 230\,$\mu$Jy/beam and 130\,$\mu$Jy/beam for the H-ATLAS
and HELMS sources respectively after combining all four spectral
windows. In this paper, we have used the calibrated visibilities as
provided in the ALMA science archive. The cleaned data used for the
image plane modelling were constructed using Briggs weighting with a
robustness parameter of -0.2 and were primary beam corrected.  Both
calibration and cleaning were carried out using version 4.3.1 of the
{\tt Common Astronomy Software Applications} package
\citep{mcmullin07}.  The image pixel scale used for the H-ATLAS and
HELMS sources was 0.02 and 0.03\,arcsec respectively.

When calculating intrinsic source properties, in addition to the
photometry obtained from our own ALMA imaging data, we have drawn from
a variety of other datasets. We have used submm photometry obtained by
the {\it Herschel} space observatory using both the Spectral and
Photometric Imaging Receiver \citep[SPIRE][]{griffin10} at the
wavelengths $250$, $350$ and $500$\,$\mu$m and the Photoconductor
Array Camera and Spectrometer \citep[PACS;][]{poglitsch10} at
wavelengths of 100 and $160$\,$\mu$m. For the H-ATLAS sources, SPIRE
and PACS photometry was taken from the H-ATLAS first data release
\citep{valiante16}.  For the HELMS sources, SPIRE fluxes were taken
from \citet[][N16 hereafter]{nayyeri16} whereas PACS fluxes were
extracted from imaging held in the {\it Herschel} Science
Archive\footnote{http://archives.esac.esa.int/hsa/whsa}. Where
available, we have also used $880\,\mu$m photometry obtained with the
Submillimeter Array (SMA) as detailed in \citet{bussmann13},
$850\,\mu$m Submillimeter Common User Bolometer Array 2 fluxes as
given in Bakx et al. (2017, in prep.) and ALMA Band 6 data
(1280\,$\mu$m) from \citet{messias14}. Finally, the source
H-ATLAS J142935.3-002836 is the Infrared Astronomical Satellite (IRAS)
source IRAS 14269-0014 for which we have taken the 60$\,\mu$m flux
density as given in the IRAS faint source catalogue \citep{moshir92}.

Table \ref{tab_lenses} lists the six systems modelled
in this paper along with their lens and source redshifts. Table 
\ref{tab_src_fluxes} gives their observed photometry.

\begin{table*}
\centering
\small
\begin{tabular}{lccccccccccc}
\hline ID & $f_{60}$ & $f_{100}$ & $f_{160}$ & $f_{250}$ & $f_{350}$ &
$f_{500}$ & $f_{850}$ & $f_{880}^{\rm SMA}$ & $f_{880}^{\rm ALMA}$ &
$f_{1280}$ \\ \hline H-ATLAS J142413.9+022303 & - & - & - & $112\pm7$
& $182\pm8$ & $193\pm8$ & $121\pm8$ & $90\pm5$ & $116\pm8$ & -
\\ H-ATLAS J142935.3-002836 & $190\pm38$ & $911\pm29$ & $1254\pm34$ &
$802\pm7$ & $438\pm7$ & $200\pm7$ & - & - & $38\pm3$ & $5.86\pm0.99$
\\ HELMS J004714.2+032454 & - & $82\pm11$ & $164\pm22$ & $312\pm6$ &
$244\pm7$ & $168\pm8$ & - & - & $49\pm5$ & - \\ HELMS J001626.0+042613
& - & $13\pm10$ & $53\pm20$ & $117\pm7$ & $151\pm6$ & $127\pm7$ & - &
- & $39\pm4$ & - \\ HELMS J004723.6+015751 & - & $104\pm15$ &
$285\pm32$ & $398\pm6$ & $320\pm6$ & $164\pm8$ & - & - & $42\pm5$ & -
\\ HELMS J001615.7+032435 & - & $23\pm11$ & $92\pm24$ & $195\pm6$ &
$221\pm6$ & $149\pm7$ & - & - & $33\pm4$ & - \\ \hline
\end{tabular}
\normalsize
\caption{Observed (i.e., lensed) source flux densities in
  mJy. Subscripts indicate the passband central wavelength in
  $\mu$m. Fluxes $f_{100}$ to $f_{500}$ inclusive are taken from the
  H-ATLAS first data release \citep{valiante16} for the two H-ATLAS
  sources. For the four HELMS sources, $f_{100}$ and $f_{160}$ are
  PACS flux densities extracted from maps acquired from the {\it
    Herschel} Science Archive and flux densities $f_{250}$ to
  $f_{500}$ are taken from \citet{nayyeri16}. Flux densities
  $f_{850}$, $f_{880}^{\rm SMA}$, $f_{880}^{\rm ALMA}$ and $f_{1280}$
  are taken from Bakx et al. (2017, in prep.), \citet{bussmann13},
  this work and \citet{messias14} respectively. Finally, $f_{60}$ is
  the 60$\,\mu$m flux taken from the IRAS faint source catalogue
  \citep{moshir92}.}
\label{tab_src_fluxes}
\end{table*}

\section{Methodology}
\label{sec_method}

In this paper, we have applied the standard image plane version of the
\citet{warren03} semi-linear inversion (SLI) lens modelling method and
a modified version which works directly in the interferometric
uv-plane on the visibility data. Both use the framework derived by
\citet{suyu06} for optimising the model Bayesian evidence.  The image
plane version adopts an implementation similar to that described by
\citet{nightingale15} which uses a randomised Voronoi tessellation in
the source plane to minimise biases in the lens model parameters. The
only differences are that here we have used k-means clustering for the
source pixels and Markov Chain Monte Carlo (MCMC) optimisation,
whereas Nightingale \& Dye used h-means clustering and MultiNest
\citep{feroz09}. The uv-plane version is described in more detail
below.

\subsection{Adapting the SLI method to visibility data}

At the heart of the SLI method lies a pixelised source plane. Using a
given lens model, an image of each pixel is formed. In the image plane
version of the method, the source surface brightness distribution for
a given lens model is determined by finding the linear superposition
of these images which best fits the observed lensed image. Adapting
this scheme to work with interferometric visibility data requires
forming a model visibility dataset for each source pixel image. The
linear combination of each model visibility dataset that best fits the
observed visibilities then recovers the source surface brightness
distribution for a given lens model, in the same manner as the image
plane SLI version.

This scheme was used recently by \citet{hezaveh16} in application to
ALMA data. In their implementation, phase calibration was included in
the modelling procedure by introducing the phase offset of each
antenna as a free parameter of the fit. In our implementation, the
sources are too faint to provide such self-calibration hence we have
instead opted to apply the phase calibration provided by external
calibrators observed throughout acquisition of our science data.

In the image plane SLI method, the rectangular matrix $f_{ij}$ holds
the fluxes of lensed image pixels $j$ for each source plane pixel $i$
assuming the source pixel has unit surface brightness. Analogously, in
the uv-plane version, the rectangular matrix $g_{ij}$ is used instead,
where each row holds the complex visibilities determined from the
lensed image of the unit surface brightness source pixel. Each row of
$g_{ij}$ therefore contains the Fourier transform of its corresponding
row in $f_{ij}$, evaluated at the same points on the uv-plane as the
observed visibilities. This is achieved by incorporating the {\tt
  MIRIAD} software package library \citep{sault95} into our
reconstruction code, but using a much streamlined version of the {\tt
  uvmodel} procedure.  The inputs to {\tt uvmodel} are the observed
visibility dataset and, in turn, the lensed images of the source plane
pixels. In this way, a model visibility dataset is created with
visibilities equal to $\sum_i \, s_i g_{ij}$ for each visibility $j$
given source pixel surface brightnesses $s_i$. With observed
complex visibilities $V_j$, the $\chi^2$ statistic is therefore
computed as \be
\label{eq_chisq}
\chi^2=\sum_{j=1}^{J} 
\frac{\sum_{i=1}^I \left| s_i g_{ij} - V_j\right|^2}{\sigma_j^2} \, ,
\ee
where the summations act over $I$ total Voronoi source pixels and $J$
visibilities and it is assumed that there is no covariance between
visibilities. We used a similar method as \citet{hezaveh16} for
determining the $1\sigma$ uncertainties, $\sigma_j$, on the
visibilities. These were computed from the rms of differences in
neighbouring visibilities grouped in the uv-plane to remove sky
contribution. Whereas Hezaveh et al. computed this for each baseline,
our computation was applied over all baselines although our analysis
excluded baselines flagged as being bad (and therefore exceptionally
noisy) by the ALMA data reduction pipeline.
The minimum $\chi^2$ solution is given by
\be
\label{eq_recon_simple}
{\rm \mathbf{s}} = {\rm \mathbf{F}}^{-1} {\rm \mathbf{v}}
\ee
where the elements of the real quantities ${\rm \mathbf{F}}$
and ${\rm \mathbf{v}}$ are respectively
\ba
{\rm F}_{ij}&=&\sum_{n=1}^{J}\frac{g^{\mathbb{R}}_{in} 
g^{\mathbb{R}}_{jn}+g^{\mathbb{I}}_{in} g^{\mathbb{I}}_{jn}}
{ \sigma_n^2} \nonumber \\
{\rm v}_i&=&\sum_{n=1}^{J}\frac{g^{\mathbb{R}}_{in} V^{\mathbb{R}}_n+
g^{\mathbb{I}}_{in} V^{\mathbb{I}}_n} {\sigma_n^2} \, .
\ea
Here, the superscripts $\mathbb{R}$ and $\mathbb{I}$ denote the real
and imaginary components respectively and the column vector ${\rm
\mathbf{s}}$ contains the real source pixel surface brightnesses.

The source is linearly regularised, introducing the real
regularisation matrix ${\rm \mathbf{H}}$ as described in
\citet{warren03}. The regularisation scheme we adopted follows that of
\citet{nightingale15}, computing the mean gradient between a given
Voronoi source pixel and its three nearest neighbours.  To find the
most probable lens model parameters, we used Markov Chain Monte Carlo
(MCMC) optimisation to maximise the Bayesian evidence derived by
\citet{suyu06}. We performed multiple MCMC runs for a range of
power-law density profile slopes which were kept fixed in each case to
help simplify parameter space. The number of source pixels was kept
fixed during optimisation and the regularisation weight was optimised
following the procedure outlined in \citet{dye08}.

\subsection{Lens model}

We used an elliptical power-law
density profile with an external shear component where necessary
to model the lenses in this work. We used the form introduced by
\citet{Ka93} which has a surface mass density, $\kappa$, 
\be
\kappa=\kappa_0\,({\tilde r}/{\rm 1kpc})^{1-\alpha} \, .
\ee
where $\kappa_0$ is the normalisation surface mass density and
$\alpha$ is the power-law index of the volume mass density profile.
Here, the elliptical radius ${\tilde r}$ is defined by ${\tilde r}^2
=x^{\prime2}+y^{\prime2}/\epsilon^2$ where $\epsilon$ is the lens
elongation (i.e., the ratio of semi-major to semi-minor axis
length). The orientation of the semi-major axis measured in a
counter-clockwise sense from north is described by the parameter
$\theta$ and the co-ordinates of the centre of the lens in the image
plane are $(x_c,y_c)$. The external shear field is characterised by
the shear strength, $\gamma$, and the shear direction angle measured
counter-clockwise from north, $\theta_\gamma$. The shear direction
angle is defined to be perpendicular to the direction of resulting
image stretch. We only incorporated external shear in the lens model
when the Bayesian evidence was improved by its inclusion. We found
that only two of the six lenses in this work needed external
shear. The total number of lens model parameters is thus eight when
shear is included and six when not.

\section{Results}
\label{sec_results}

\begin{table*}
\centering
\small
\begin{tabular}{llcrcccrc}
\hline
ID & $\kappa_0$ & $(x_c,y_c)$ (arcsec) & $\theta({\rm deg})$ & $\epsilon$ & $\alpha$ & $\gamma$ & $\theta_\gamma({\rm deg})$ & $\theta_E({\rm arcsec})$\\
\hline
{\em Image plane} & & & & & & & & \\
\hline
H-ATLAS J142413.9 & $0.59\pm0.01$ & $(0.18\pm 0.01,0.68\pm0.01)$ & $84\pm2$ & $1.07\pm0.02$ & $2.03\pm0.04$ &  &  & $0.97\pm0.04$ \\
H-ATLAS J142935.3 & $0.44\pm0.01$ & $(1.60\pm 0.01,0.62\pm 0.01)$ & $124\pm1$ & $1.33\pm0.02$ & $1.82\pm0.05$ &  &  & $0.71\pm0.03$\\
HELMS J004714.2   & $0.50\pm0.01$ & $(1.56\pm 0.02,2.34\pm 0.03)$ & $94\pm2$ & $1.25\pm0.02$ & $1.96\pm0.04$ &  &  & $0.59\pm0.03$\\
HELMS J001626.0   & $0.56\pm0.01$ & $(2.88\pm 0.02,1.67\pm 0.02)$ & $36\pm1$ & $1.37\pm0.03$ & $2.14\pm0.06$ &  &  & $0.98\pm0.07$\\
HELMS J004723.6   & $1.18\pm0.02$ & $(2.52\pm 0.02,-0.60\pm 0.02)$ & $178\pm2$ & $1.18\pm0.01$ & $1.87\pm0.04$ & $0.09\pm0.01$ & $167\pm2$ & $2.16\pm0.10$\\
HELMS J001615.7   & $2.21\pm0.04$ & $(0.12\pm 0.05,-0.96\pm 0.07)$ & $18\pm2$ & $1.41\pm0.02$ & $2.00\pm0.07$ & $0.13\pm0.01$ & $55\pm2$ & $2.79\pm0.20$\\
\hline
{\em Visibility plane} & & & & & & & & \\
\hline
H-ATLAS J142413.9 & $0.59\pm0.01$ & $(0.18\pm 0.01,0.68\pm0.01)$ & $85\pm2$ & $1.07\pm0.01$ & $2.06\pm0.04$ &  &  & $0.97\pm0.04$ \\
H-ATLAS J142935.3 & $0.43\pm0.01$ & $(1.60\pm 0.01,0.61\pm 0.01)$ & $125\pm1$ & $1.35\pm0.02$ & $1.79\pm0.05$ &  &  & $0.70\pm0.03$\\
HELMS J004714.2   & $0.50\pm0.01$ & $(1.55\pm 0.02,2.34\pm 0.03)$ & $93\pm2$ & $1.24\pm0.02$ & $1.91\pm0.05$ &  &  & 0.$58\pm0.03$\\
HELMS J001626.0  & $0.58\pm0.01$ & $(2.89\pm 0.02,1.66\pm 0.02)$ & $36\pm1$ & $1.38\pm0.03$ & $2.18\pm0.06$ &  &  & $0.98\pm0.07$\\
HELMS J004723.6   & $1.18\pm0.03$ & $(2.51\pm 0.02,-0.60\pm 0.02)$ & $178\pm2$ & $1.20\pm0.02$ & $1.89\pm0.06$ & $0.08\pm0.01$ & $161\pm2$ & $2.08\pm0.12$\\
HELMS J001615.7   & $2.00\pm0.07$ & $(0.11\pm 0.05,-0.94\pm 0.06)$ & $18\pm2$ & $1.42\pm0.02$ & $1.90\pm0.05$ & $0.10\pm0.01$ & $53\pm2$ & $2.96\pm0.16$\\
\hline
\end{tabular}
\normalsize
\caption{Lens model parameters. The top half of the table gives the
  parameters obtained from the image plane analysis and the bottom
  half gives those from the visibility plane analysis. Only HELMS
  J004723.6+015751 and HELMS J001615.7+032435 showed significant
  improvement in the fit when external shear was included in the lens
  model, hence the remaining four were modelled without it. Parameters
  are: lens normalisation, $\kappa_0$ in units of
  $10^{10}M_\odot\,{\rm kpc}^{-2}$; co-ordinates of the lens model
  centroid with respect to the phase-tracking centre of observations
  (west and north correspond to positive $x_c$ and $y_c$
  respectively); lens semi-major axis orientation, $\theta$, measured
  counter-clockwise from north; lens semi-major to semi-minor axis
  ratio, $\epsilon$; logarithmic slope of the power-law density
  profile, $\alpha$; external shear strength, $\gamma$; shear
  direction angle, $\theta_\gamma$, measured counter-clockwise from
  north; Einstein radius, $\theta_E$.}
\label{tab_lenspars}
\end{table*}

Figure \ref{recon} shows the model reconstructions of each of the six
lenses using both the image plane and visibility plane methods.  It is
apparent from the figure that whilst there are differences in the
reconstructed sources between both methods, these are quite
subtle. The variation in source plane pixelisation between image plane
and uv-plane reconstructions likely accounts for a significant amount
of this variation; the largest difference in morphology is seen in the
case of H-ATLAS J142935.3-002836 but, owing to the random nature of
the k-means clustering, this source also possesses the largest
differences in source pixelisation. An anticipated tendency of the
image plane method to reproduce possible artifacts arising from
transformation from the visibility plane or cleaning procedure has not
manifested itself in the reconstructions.  Faint source features seen
in each lens system are commonly reconstructed with both methods,
giving an indication of their robustness. Additionally, the fact that
the optimal regularisation weight may differ between the image and
visibility plane due to correlated image plane pixels appears to have
had little consequence\footnote{We adopted a uniform noise map for the
  image plane modelling, neglecting correlations between image pixels
  although we found that varying the pixel scale produced no
  significant changes in the reconstruction.}, although this effect
may be at least partly responsible for the differences seen between
some residual plots. (For example, H-ATLAS J142935.3-002836 and HELMS
J001626.0+042613 show significant residuals at the location of image
peaks in the image plane reconstruction compared to the uv-plane
reconstruction.) The strongest features identified in the residual plots,
such as those of H-ATLAS J142935.3-002836 and HELMS J004714.2+032454,
have a significance of $\sim 2.5\sigma$.

Figure \ref{recon} also shows the dirty beam maps for each lens
system.  The strongest sidelobes occur in the HELMS beams
approximately 1\,arcsec east and west of the central beam
component. These sidelobes each contain 6 per cent of the flux
contained in the main beam component. To assess the impact that such
sidelobes might have on the reconstructions, we carried out a simple
test whereby we reconstructed the cleaned image of HELMS
J001626.0+042613 with the dirty beam and the model beam. The resulting
reconstructions showed differences in the source and model images
which were only at the level of a few per cent, smaller than the
differences between uv-plane and image-plane reconstructions. We
therefore conclude that beam sidelobes in the current data play a
negligible role.

\begin{figure*}
\epsfxsize=15.8cm
{\hfill
\epsfbox{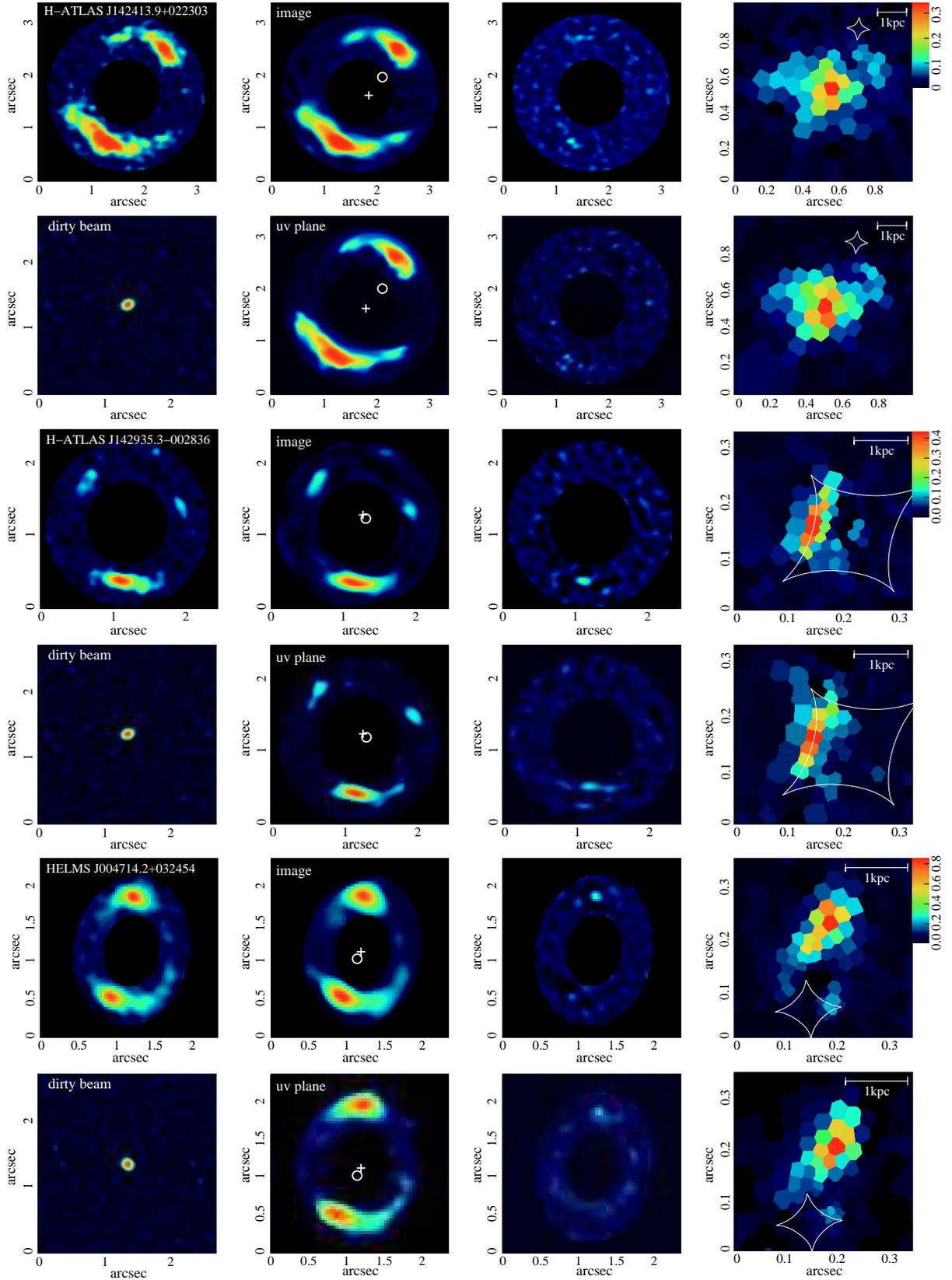}
\hfill}
\epsfverbosetrue
\caption{Lens reconstructions. Each system is shown in pairs of rows,
  the cleaned ALMA image and the dirty beam being shown in the top
  left-most and bottom left-most panels respectively. The middle-left,
  middle right and right-most columns show the image of the
  reconstructed source (the model image -- the white cross and
    white circle shows the source plane centre and lens model centroid
    respectively), the cleaned image minus the model image and the
  reconstructed source respectively, the top row showing the image
  plane reconstruction and the bottom row showing the visibility plane
  reconstruction.  The reconstructed source plots show the caustic
  (white lines). The colour scale gives the surface brightness at
    880$\,\mu$m in Jy\,arcsec$^{-2}$ for source and image plots. All 
   residuals are $<3\sigma$.}
\label{recon}
\end{figure*}

\begin{figure*}
\epsfxsize=15.8cm
{\hfill
\epsfbox{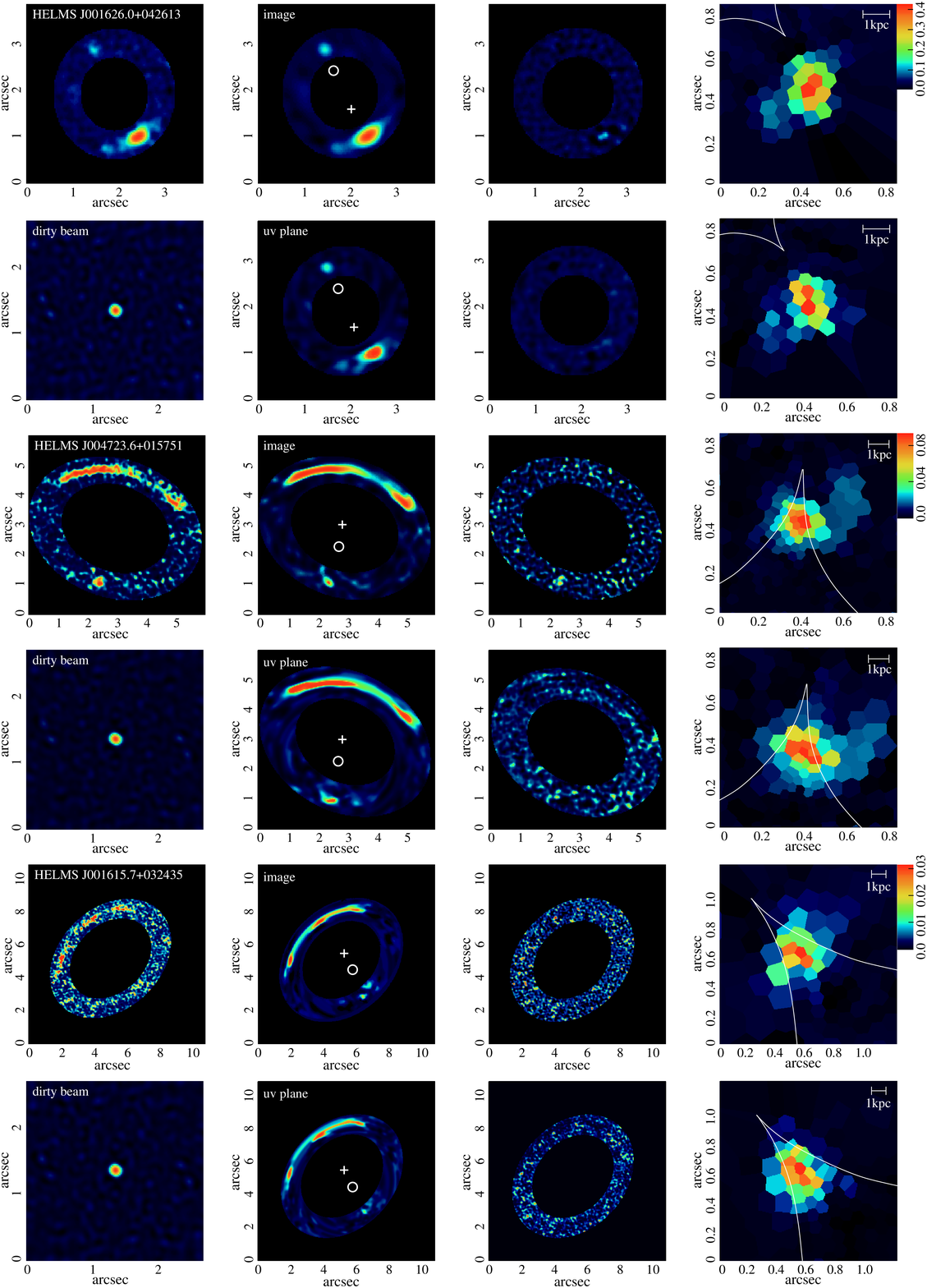}
\hfill}
\epsfverbosetrue
\contcaption{Lens reconstructions. }
\label{recon2}
\end{figure*}

The lens model parameters recovered for each of the six lenses using
the image plane and visibility plane methods are given in table
\ref{tab_lenspars}. On the whole, there is good agreement between the
parameters obtained using the two methods, although there are
mildly significant differences in the case of HELMS J001615.7+032435.
However, this system has the lowest signal to noise ratio and the lack
of detection of a counter image introduces additional uncertainty.

\begin{figure*}
\epsfxsize=18cm
{\hfill
\epsfbox{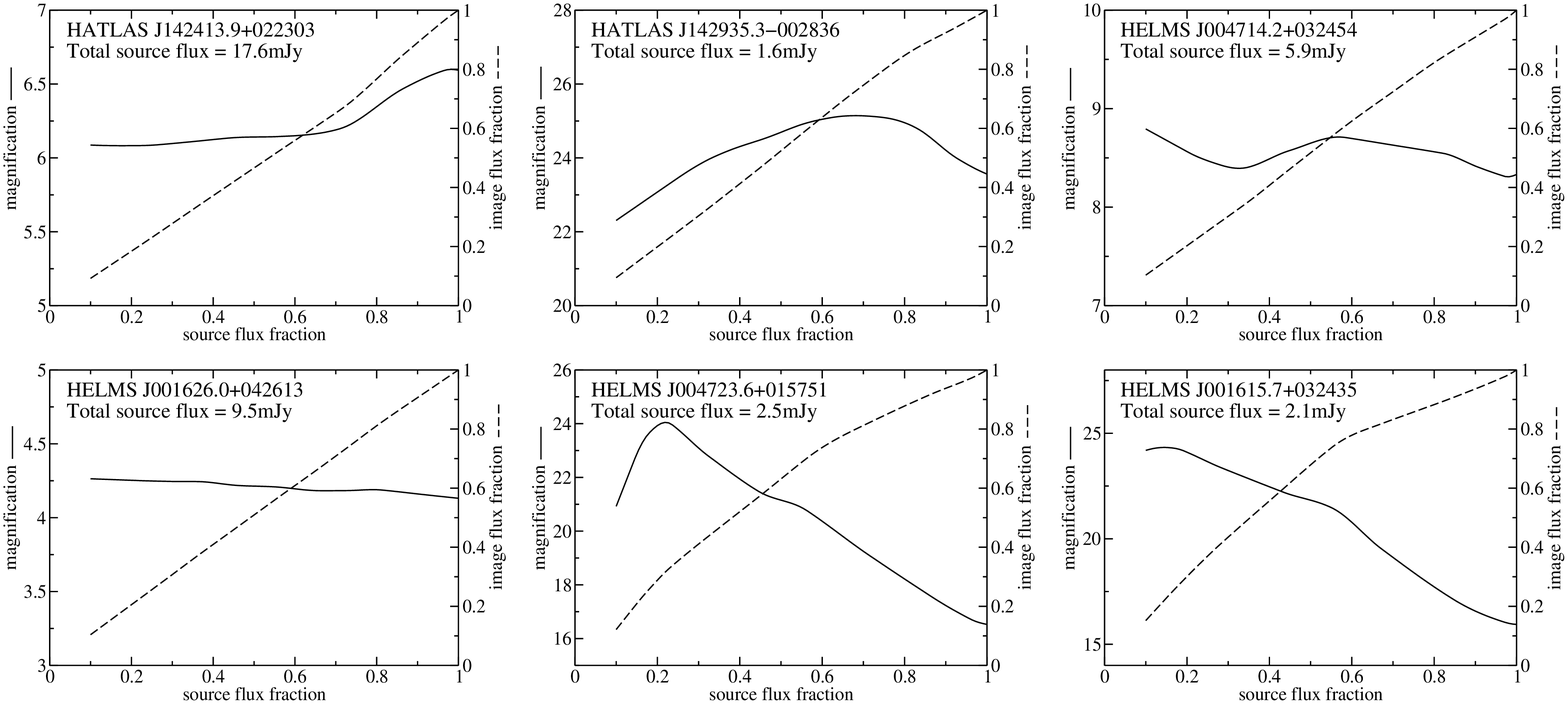}
\hfill}
\epsfverbosetrue
\caption{Magnification profile plots of image plane reconstructions.
  Each panel shows how magnification (solid line) and image flux
  density fraction (dashed line) varies as a function of the fraction
  of total source flux density above a surface brightness threshold
  (see main text for details). Magnification profiles have been
  averaged over 100 realisations of the source plane pixelisation for
  the best-fit lens model. The plot gives an indication of the extent
  to which the computed magnification varies with source surface
  brightness as would be reached by different interferometer
  configurations. The largest variation in magnification is seen for
  HELMS J004723.6+015751 and HELMS J001615.7+032435 since both
  have sources located in the vicinity of a lensing caustic cusp.}
\label{mag_plots}
\end{figure*}

Figure \ref{mag_plots} shows how source magnification varies as a
fraction of ranked source surface brightness. We took the best fit
lens model for each system (determined from the image plane modelling
although the results are very similar from the uv-plane modelling --
see table \ref{tab_src_props}) and computed the average source
magnification factor of 100 different source plane pixelisations. This
was computed for different fractions of the total source flux density
by working down a list of source pixels ranked by flux density
(i.e. the product of source pixel area and reconstructed surface
brightness). The plots show how sensitive the inferred magnification
is to different interferometric configurations which probe different
scales and surface brightness limits. The two systems HELMS
J004723.6+015751 and HELMS J001615.7+032435 exhibit the largest
variation in magnification since their sources are located in the
vicinity of a caustic cusp where magnification gradients are
significantly stronger.

\subsection{Intrinsic source properties}
\label{sec_src_props}

We have computed intrinsic properties of the background sources in
each lens system. To do this, we de-magnified the available submm
photometry (see table \ref{tab_src_fluxes}) by the total source
magnification factors derived from the image plane reconstructions,
$\mu^{\rm img}_{\rm tot}$, as given in table
\ref{tab_src_props}. These are consistent with the magnifications from
the uv-plane reconstructions in the sense that all differences in
magnification propagate to differences in intrinsic source properties
that are significantly smaller than the uncertainties arising from the
SED fitting. Using the source redshifts given in table
\ref{tab_lenses}, we then fitted the rest-frame photometry with both a
single temperature optically thick spectral energy distribution (SED)
and a dual temperature optically thin SED. This SED choice gives an
estimate of the upper and lower values in the range of possible dust
masses, which we computed using the method outlined in
\citet{dunne11}. Here, we used the observed ALMA 880\,$\mu$m flux
density and a dust mass absorption coefficient computed by
extrapolating the 850\,$\mu$m value of
$\kappa_{850}=0.077$\,m$^2$kg$^{-1}$ \citep{james02} to the rest-frame
wavelength corresponding to the observer-frame wavelength of
880\,$\mu$m \citep[see][for more details]{dunne00}.  Computing dust
masses in this way minimises the propagation of errors in dust
temperature.

When fitting the optically thin SED, the temperature and normalisation
of both components were varied. For the optically thick SED,
temperature, normalisation and the opacity at 100\,$\mu$m,
$\tau_{100}$, were varied in the fit. In all cases, the emissivity
index was fixed to 2.0 \citep[see, for example][]{smith13}. The best
fit SED parameters and the corresponding de-magnified luminosity of
the source computed by integrating the best fit optically thin SED
from 3-1100\,$\mu$m are given in table \ref{tab_src_props}.  Finally,
we computed the star formation rate of the source with the conversion
from luminosity given by \citet{kennicutt12} which uses a Kroupa
\citep{kroupa01} initial mass function (IMF).

%LD's two-component fits with thin and thick dust and beta=2.0:
\begin{table*}
\centering
\small
\begin{tabular}{lcccccccccc}
\hline
ID & $\mu^{\rm img}_{\rm tot}$ & $\mu^{\rm uv}_{\rm tot}$ & $M_d^{\rm thick}$ & $M_d^{\rm thin}$ & $T^{\rm thick}/K$ & $T^{\rm thin}/K$ & $\tau_{100}$ & $L_{\rm FIR}$ & $M_{\rm gas}$& SFR ($M_\odot$/yr) \\
\hline
H-ATLAS J142413.9 & $6.6\pm0.5$ & $6.4\pm0.5$   & 8.7 & 9.7 & 59 & 41 / 21 & 5.8 & $13.2\pm0.1$ & $11.8\pm0.1$ & $2200 \pm 500$\\
H-ATLAS J142935.3 & $23.6\pm1.3$ & $22.3\pm1.3$ & 7.9 & 8.2 & 70 & 45 / 26 & 4.4 & $12.3\pm0.1$ & $10.7\pm0.1$ & $330 \pm 80$\\
HELMS J004714.2   & $8.3\pm0.6$  & $8.7\pm0.6$  & 8.7 & 9.2 & 43 & 51 / 22 & 9.2 & $12.2\pm0.1$ & $11.3\pm0.1$ & $220 \pm 60$\\
HELMS J001626.0   & $4.1\pm0.3$  & $4.3\pm0.3$  & 8.8 & 9.3 & 48 & 57 / 27 & 4.4 & $12.8\pm0.1$ & $11.5\pm0.1$ & $980 \pm 240$\\
HELMS J004723.6   & $16.5\pm1.0$ & $15.2\pm1.0$ & 8.2 & 8.7 & 52 & 48 / 26 & 5.2 & $12.2\pm0.1$ & $10.9\pm0.1$ & $230 \pm 60$\\
HELMS J001615.7   & $15.9\pm1.0$ & $17.1\pm1.0$ & 7.9 & 8.5 & 58 & 72 / 34 & 2.4 & $12.5\pm0.1$ & $10.7\pm0.1$ & $480 \pm 100$\\
\hline
\end{tabular}
\normalsize
\caption{Intrinsic source properties. Columns are the total source
  magnification computed using the image plane method and uv-plane
  method, $\mu^{\rm img}_{\rm tot}$ and $\mu^{\rm uv}_{\rm tot}$
  respectively, dust mass assuming a single temperature optically
  thick SED, $M_d^{\rm thick}$, dust mass assuming a dual temperature
  optically thin SED, $M_d^{\rm thin}$, temperature of the optically
  thick SED, $T^{\rm thick}$, temperatures of the optically thin SED,
  $T^{\rm thin}/K$, the opacity at $100\,\mu$m for the optically thick
  SED, $\tau_{100}$, de-magnified luminosity (computed as the integral
  of the best fit SED from 3 to 1100\,$\mu$m using the optically thin
  SED), $L_{\rm FIR}$, H$_2$ gas mass calculated using the scaling
  relation of \citet{hughes17}, $M_{\rm gas}$, and star formation rate
  (SFR) scaled from $L_{\rm FIR}$ using the prescription given by
  \citet{kennicutt12} with a Kroupa IMF. Dust masses are expressed as
  $\log_{10}(M_d/M_\odot)$, gas masses as $\log_{10}(M_{\rm
    gas}/M_\odot)$ and the luminosity values are $\log(L_{\rm
    FIR}/L_\odot)$.}
\label{tab_src_props}
\end{table*}

\subsubsection{Object notes}

\begin{figure}
\epsfxsize=8.5cm
{\hfill
\epsfbox{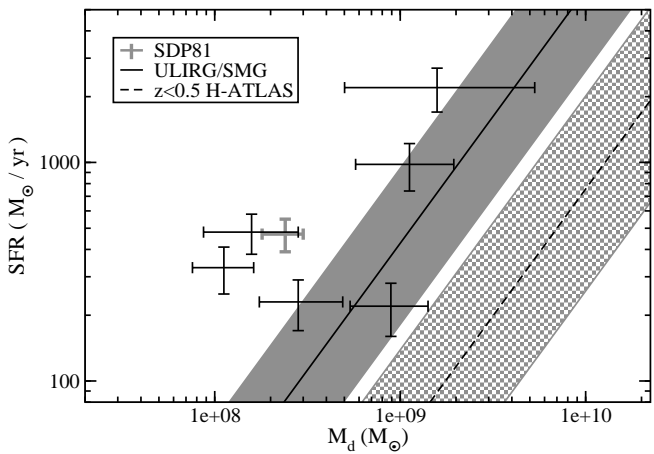}
\hfill}
\epsfverbosetrue
\caption{Star formation rate \citep[determined using the method of]
  []{kennicutt12} plotted against dust mass for the six lensed
  sources. For each source, the range in dust mass spanned by
  $M_d^{\rm thick}$ and $M_d^{\rm thin}$ is plotted, with
  uncertainties in SFR indicated at the midpoint. Also plotted are the
  empirical relationships between SFR and $M_d$ determined by
  \citet{rowlands14} for high redshift SMGs and low redshift ULIRGs
  (solid line with 1$\sigma$ spread indicated by the solid grey shaded
  region) and the population of $z<0.5$ galaxies detected in H-ATLAS
  (dashed line with 1$\sigma$ spread indicated by the perforated grey
  shaded region). The thick grey cross locates SDP.81 as determined by
  \citet{dye15}. One interpretation of this plot is that the majority
  of lensed sources in this paper have higher dense molecular gas
  fractions than the average ULIRG/SMG (see Section \ref{sec_summary}
  for more discussion).}
\label{SFR_Md}
\end{figure}

\begin{itemize}
\setlength\itemsep{1em}

\item[ ] {\em H-ATLAS J142413.9+022303} - 
% GAMA15-1
Keck K-band imaging of this system \citep[see][]{calanog14} reveals
two compact galaxies interior to the Einstein ring, each consistent
with an early-type morphology. Follow-up spectroscopy by \citet[][B12
  hereafter]{bussmann12} gives a redshift of $z=0.595$ but due to lack
of spatial resolution, it is unclear if this corresponds to solely the
brighter primary galaxy or whether both galaxies have the same
redshift.  In this work, we have used a single power-law profile,
finding that this gives a perfectly acceptable fit to the data. The
lens profile centre, which is a free parameter of the fit, aligns
within 0.05\,arcsec of the centre of the brighter of the two
galaxies. Adding a second mass to the lens model does not provide a
significant improvement to the fit and makes a negligible difference
to the inferred intrinsic source properties reported herein.

B12 found that a source model comprising two sersic profiles gives a
significantly better fit than a single sersic profile source model. At
a qualitative level, this is consistent with the irregular morphology
of the reconstructed source we have obtained in the current work. B12
also estimated the de-magnified luminosity of the CO(1-0) line emitted
by the source and found this to be a factor of 2.4 greater than that
inferred from the line dispersion \citep[which correlates with line
  luminosity; see, for example][]{harris12}. This discrepancy is
significantly lessened to 1.4 using our magnification factor which is
80 per cent higher than that determined by B12.

The lensed source in this system has a very high star formation rate
(SFR) of 2200\,$M_\odot$/yr (see below for more discussion). This
compares to the value of $\simeq 5000\,M_\odot$/yr reported by
\citet{bussmann13}, although this becomes $\simeq 2800\,M_\odot$/yr
using our magnification factor instead.

\begin{figure*}
\epsfxsize=18cm
{\hfill
\epsfbox{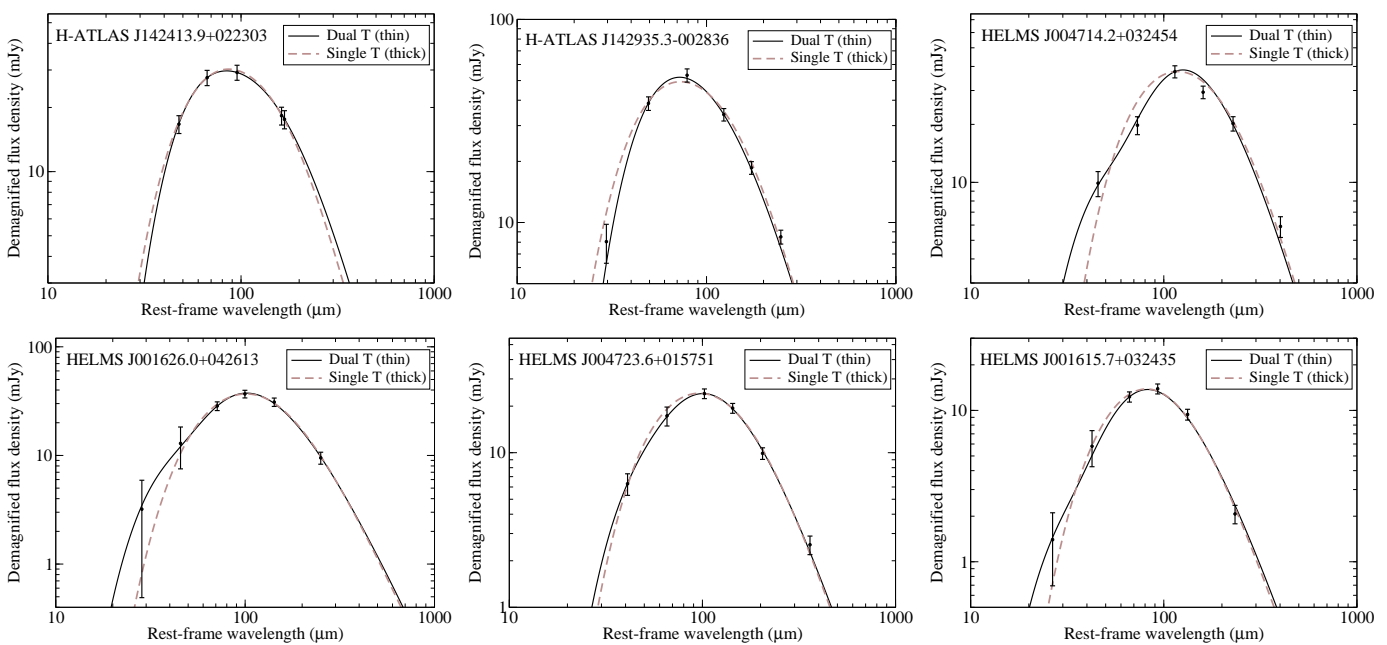}
\hfill}
\epsfverbosetrue
\caption{SEDs of the lensed sources. Each plot shows the
  two-temperature optically thin fit (continuous black line) and the
  single-temperature optically thick fit (dashed grey line). The
  measured photometry shown by the data points in the plots are
  de-magnified using the total magnifications, $\mu^{\rm img}_{\rm
    tot}$, given in table \ref{tab_src_props}.}
\label{SEDs}
\end{figure*}

\item[ ] {\em H-ATLAS J142935.3-002836} - 
% GAMA15-3
This lens system has been previously investigated in detail by
\citet[][M14 hereafter]{messias14} who analysed a broad range of
multi-wavelength imaging, including ALMA Band 3 and Band 6 data (with
central wavelengths of 3.1\,mm and 1.3\,mm respectively and maximum
resolutions of 1.4\,arcsec and 0.6\,arcsec respectively). Optical
imaging acquired with the Keck telescope \citep[see][]{calanog14}
indicates that the lens is an edge-on spiral and optical spectroscopy
by M14 from the Gemini-South telescope gives a lens redshift of 0.218.

The power-law lens model determined by M14 using image plane modelling
of their submm/mm data has parameters $\kappa_0=(0.40 \pm 0.01) \times
10^{10}M_\odot\,{\rm kpc}^{-2}$, $\alpha=2.08\pm0.08$,
$\epsilon=1.46\pm0.04$, $\theta=136\pm1\,{\rm deg}$ and
$\theta_E=0.62\pm0.08$\,arcsec compared to the parameters
$\kappa_0=(0.43\pm0.01) \times 10^{10}M_\odot\,{\rm kpc}^{-2}$,
$\alpha=1.79\pm0.05$, $\epsilon=1.35\pm0.02$, $\theta=125\pm 1\,{\rm deg}$
and $\theta_E=0.70\pm0.03$\,arcsec obtained directly from our much
higher resolution ALMA visibility data. Whilst the models are similar,
there are some significant discrepancies in certain parameters. One
likely cause of this might stem from degeneracies between the triplet
$\kappa_0$, $\alpha$ and $\epsilon$ which can give rise to substantial
differences if any systematics are present \citep[for example, arising
  from the fixed source plane grid used in the modelling method of
  M14; see][for more details]{nightingale15}.

Our reconstructed ALMA Band 7 source has the same linear structure as
that found by M14 in the submm/mm wavebands, aligned with
approximately the same orientation along the lens fold caustic.
Regarding the source magnification factor, our value of 24 is
consistent with the values quoted in M14\footnote{In M14,
magnifications were computed over different fractions of the source
plane area containing 10, 50 and 100 per cent of the total source
plane flux. M14 computed a 50 per cent magnification of 14 and a 10
per cent magnification of 26. To be consistent with the definition
used by M14 would require a source plane fraction somewhere between
these two values.}.  In our reconstruction, there is a hint of
morphological disturbance at the southern end of the source. This is
exactly where M14 find that a second optically detected source
intersects in what they interpret as a possible merger.

This source has an extremely high SFR to dust mass ratio, the highest
in our sample.  The source lies $>3\sigma$ away from the mean in
the distribution of SFR to dust mass ratios of high redshift submm
galaxies (SMGs) and lower redshift ULIRGs determined by
\citet{rowlands14} as Figure \ref{SFR_Md} shows.

\item[ ]{\em HELMS J004714.2+032454} -
%HELMS-1
This is a double image system which is very well fit with a single
power-law density profile and no external shear. The source exhibits a
long faint structure extending to the south-east and this is readily
seen in the lensed image.

The SPIRE and ALMA photometry alone continues to rise towards shorter
wavelengths, the peak of the SED being constrained purely by the PACS
photometry. The relatively high 100$\,\mu$m PACS flux is suggestive of
a warmer dust component and this is reflected in a significantly
better fit by the dual temperature SED compared to the single
temperature template, although both SEDs give a comparable dust mass.

De-magnifying the far-IR luminosity given in N16 using our
magnification factor of 8.3 gives $\log (L_{\rm
  FIR}/L_\odot)=12.1\pm0.1$, slightly less than our determination but
consistent within the uncertainties. The luminosity implies a star
formation rate of $\simeq 220\pm60$\,$M_\odot$/yr. Given its dust mass
range of $10^{8.7}-10^{9.2}$ $M_\odot$, this places the source
somewhere between having the characteristics of a high redshift SMG or
lower redshift ULIRG and the bulk population of $z<0.5$ galaxies
detected in H-ATLAS, according to \citet{rowlands14}.

\item[ ] {\em HELMS J001626.0+042613} -
%HELMS-2
This double image system is well described by an isolated power-law
density profile and a relatively compact source. Both reconstruction
methods suggest faint extended source structure but this only
contributes a few per cent of the main source flux.  The system has
the lowest magnification factor in our sample of only
$4.1\pm0.3$. 

The peak of the source SED in this system is well bounded by the ALMA
and SPIRE photometry giving robust temperature estimates. In the dual
temperature SED, the warm component makes a larger contribution to the
total dust mass than the other five sources but this is not well
constrained owing to uncertainties in the shorter wavelength PACS
photometry.  The de-magnified source luminosity is $\log (L_{\rm
  FIR}/L_\odot)=12.7\pm0.1$ which agrees with the value quoted by
N16. The $z=2.51$ source has a high SFR of $980$\,$M_\odot$/yr and its
SFR to dust mass ratio is consistent with a typical SMG/ULIRG as
indicated in Figure \ref{SFR_Md}.

\item[ ] {\em HELMS J004723.6+015751} -
%HELMS-5
This system is one of two in our sample which require external
shear in the lens model, consistent with the location of a smaller
external galaxy 10\,arcsec to the south. The source shows a
compact, relatively featureless morphology with the hint of an
extended structure to the north west.

The SPIRE and ALMA photometry of the source on their own indicate that
the peak of the SED lies in the vicinity of the shortest wavelength
data point at 250\,$\mu$m. This is borne out by the inclusion of
PACS photometry. As a result, the fitted dual temperature SED implies a
dominant mass of cold dust at 26\,K. The intrinsic source luminosity
of $\log (L_{\rm FIR}/L_\odot)=12.2$ is in agreement with that
measured by N16. The SFR of $230\pm60$\,$M_\odot$/yr for this $z=1.44$
source compared with its relatively low dust mass places it in the
upper envelope of SFR to dust mass ratios spanned by SMGs and ULIRGs
according to \citet{rowlands14}.

\item[ ] {\em HELMS J001615.7+032435} -
%HELMS-6
The relatively low image signal-to-noise ratio in this cusp-caustic
configuration lens results in an undetected counter-image which
increases the modelling uncertainty for this system. Nevertheless, the
most probable lens model is one with a significant external
shear. This is consistent with several smaller nearby galaxies, mainly
to the north-east, with colours similar to the lens which is,
in turn, consistent with the larger Einstein radius of a group-scale
lens.

In light of this, we attempted a lens model that includes external
convergence provided by a singular isothermal ellipsoid (SIE) mass
model. The best fit model we found places the SIE to the north-east
with the result that the required external shear is reduced by
approximately 30 per cent and the normalisation of the primary lens,
$\kappa_0$, is lowered by approximately 20 per cent. The magnification
is also reduced by approximately 30 per cent. However, the model is
less favoured by the Bayesian evidence and there is a tendency for it
to produce a brighter counter image which would have been detected in
the ALMA data. The location and normalisation of the external SIE is,
as expected, degenerate with the normalisation and shear of the
primary lens. Further observations of the lensing galaxies are
required to better characterise the lens model.

The ALMA and SPIRE photometry of the source in this lens system
prefers an optically thick single temperature SED. However, with the
inclusion of PACS fluxes, a marginally improved fit is obtained with a
second weak but quite hot dust component, although the improvement in
the fit is not significant given the additional SED parameters.  The
source has a luminosity of $\log (L_{\rm FIR}/L_\odot)=12.5$ which
agrees with that of N16 who used an optically thin single component 
SED.  The SFR to dust mass ratio of this source is extremely high,
placing it nearly $3\,\sigma$ above the mean in the distribution of
ratios measured in the SMG/ULIRG population.

\end{itemize}

\section{Summary and Discussion}
\label{sec_summary}

We have modelled ALMA imaging data of six strong galaxy-galaxy
gravitational lens systems originally detected by the {\it Herschel}
Space Observatory.  For each lens system, we have carried out
modelling of both the cleaned image data and the visibility data
directly. We find only minor differences in the reconstructed source
morphologies between the two methods. The expectation is that such
differences will become more prominent as coverage of the uv-plane
becomes more sparse, not least because this will generally lead to
larger-scale image pixel covariances from beam sidelobes which are not
included in the cleaned data. In \citet{dye15}, modelling of the
cleaned image was advocated on the basis that the uv-plane was very
well sampled in that particular case and because image plane modelling
is substantially more computationally efficient than uv-plane
modelling generally.  In the present work, the uv-plane is less well
sampled in comparison and hence the decrease in efficiency by
modelling the visibility data is less severe. Nevertheless, image plane
modelling is still at least an order of magnitude quicker than uv-plane
modelling and gives very similar results.

In our fitting of a smooth power-law mass density profile, we have
found that the lenses are all close to isothermal and that the
recovered model parameters are in broad agreement between both
methods. However, one system with particularly poor signal to noise
shows mildly significant discrepancies in the slope and normalisation
of the power-law profile, although these two parameters are typically
quite degenerate. A more exhaustive investigation into the origin,
prevalence and strength of such discrepancies along with differences
in the reconstructed source is left for future study.

We have used the lens magnification factors obtained from the
modelling to demagnify the submm source photometry. Fitting rest-frame
SEDs to this photometry, we have determined the dust temperature, dust
mass, luminosity and inferred star formation rate of the lensed
sources. Using both an optically thick single-temperature SED and an
optically thin SED with two temperature components has allowed an
estimate of the range of dust mass possible for each source.  Taking
the mid-point of this range in each case, we find that five of the six
sources have a ratio of star formation rate to dust mass which is in
excess of the mean ratio of the SMG/ULIRG population as determined by
\citet{rowlands14}.

The extent of this excess is shown in Figure \ref{SFR_Md} which plots
the SFR obtained by scaling the far-IR luminosity using the relation
given by \citet{kennicutt12} against dust mass. The figure shows that
two of the sources in our sample are at least as extreme as the
H-ATLAS lensed source SDP.81 investigated by \citet{dye15}. These lie in
the upper envelope of the distribution of SFR-to-dust mass measured by
\citet{rowlands14}.  Since our computed SFR is simply a scaled version
of far-IR luminosity, the underlying fact is that these sources have a
high luminosity for the quantity of gas available for star
formation. This is often an indication that a component of the
source's luminosity comes from an active galactic nucleus but we are
unable to comment further on this possibility without additional
observations.

If we convert the rest-frame 850$\mu$m flux density of our sources to
H$_2$ gas mass (see table \ref{tab_src_props}) using the empirical
scaling relation given by \citet{hughes17}, we find that the five
sources located above the Rowlands et al. SFR-to-dust mass
relationship also lie on or above the mean relationship between SFR
and H$_2$ gas mass determined by \citet{scoville16}. If dust is indeed
an accurate tracer of molecular gas as these scaling relationships
suggest, then the implication is that these sources possess a higher
star formation efficiency (SFE).  Treating the range in dust mass for
each source as a 1-sigma error and fitting a line parallel to the
SMG/ULIRG relationship in Figure \ref{SFR_Md} to the mid-point of the
dust mass range for all six sources, the increase in SFE is a factor
of 5 relative to that implied by the SMG/ULIRG relationship of
Rowlands et al. and a factor of 40 relative to $z<0.5$ H-ATLAS
galaxies.

An alternative explanation to the SFR-to-dust mass offset being the
result of an enhanced SFE could be that the gas-to-dust ratio in these
sources is higher. Similarly, the results would be explained if the
dust mass opacity coefficient were lower by the factors mentioned
above. Both of these possibilities seem to disagree with measurements
of gas mass from CO detections at low and high redshift \citep[see,
  for example,][]{dunne01,magdis12,rowlands14,
  scoville14,scoville16,grossi16,hughes17}. These studies indicate a
tight correlation between CO line intensity and 850$\,\mu$m
luminosity, thereby implying a constant H$_2$ gas-to-dust mass ratio.
However, a caveat is that this assumes a fixed value of the ratio of
H$_2$ surface gas mass density to CO line intensity, $\alpha_{\rm
  CO}$. \citet{sandstrom13} find a weak dependence of $\alpha_{\rm
  CO}$ on metallicity in local galaxies, such that lower metallicity
tends to correspond to higher values of $\alpha_{\rm CO}$. If this
holds in high redshift SMGs, whilst a lower metallicity would not
affect the CO-to-dust ratio, the ratio of H$_2$ gas-to-dust would be
increased, leading to an enhanced SFR-to-dust mass ratio.

An additional point to note is that interpreting a higher SFR to gas
mass ratio as a higher SFE when the total molecular gas mass is used
assumes that star formation occurs throughout the full extent of
molecular gas. Determinations of dense molecular gas mass traced by
HCN emission show a correlation between far-IR luminosity and HCN line
intensity that is much tighter than the correlation between HCN and CO
line intensity \citep[see for example,][]{gao04,privon15}. SFR
therefore appears to depend on dense molecular gas mass rather than
total molecular gas mass traced by CO. In light of this, and assuming
universal star formation physics, a more probable interpretation of
the high SFR to gas mass ratios we find is that the sources in our
sample have a significantly higher dense molecular gas mass
fraction. This conclusion was also reached by \citet{oteo17} who carried
out a similar analysis of two H-ATLAS lensed sources.

\citet{pap12} provide evidence to suggest that high density molecular
gas is more prevalent in galaxy mergers than quiescently forming
systems and that its fraction can be used to determine the mode of
star formation.  Inspection of the reconstructed morphologies (Figure
\ref{recon}) of the two sources in our sample with extreme SFR to gas
mass ratios (i.e., HELMS J001615.7+032435 and H-ATLAS
J142935.3-002836) does indeed reveal signs of disturbed morphology,
but no more so than others in the sample. Nevertheless, increasing the
number of gravitational lens reconstructions of such systems with high
magnification factors offers the ability to further investigate such
hypotheses. This becomes especially true with the inclusion of source
kinematics measured via molecular lines.

\section*{Acknowledgements}

SD acknowledges support from the UK STFC Ernest Rutherford Fellowship
scheme. LD acknowledges funding from the European Research Council
Advanced Investigator grant COSMICISM and the ERC Consolidator grant
CosmicDust.  MN acknowledges financial support from the European
Union's Horizon 2020 research and innovation programme under the Marie
Sk{\l}odowska-Curie grant agreement No 707601. MJM acknowledges the
support of the National Science Centre, Poland through the POLONEZ
grant 2015/19/P/ST9/04010.  LM acknowledges support from the South
African National Research Foundation through the South African
Research Chairs Initiative. This project has received funding from the
European Union's Horizon 2020 research and innovation programme under
the Marie Sk{\l}odowska-Curie grant agreement No. 665778. Some of the
spectroscopic redshifts reported in this paper were obtained with the
Southern African Large Telescope (SALT) under proposal
2015-2-MLT-006. This paper makes use of the following ALMA data:
ADS/JAO.ALMA\#2013.1.00358.S. ALMA is a partnership of ESO
(representing its member states), NSF (USA) and NINS (Japan), together
with NRC (Canada) and NSC and ASIAA (Taiwan) and KASI (Republic of
Korea), in cooperation with the Republic of Chile. The Joint ALMA
Observatory is operated by ESO, AUI/NRAO and NAOJ. This research has
made use of the NASA/IPAC Infrared Science Archive, which is operated
by the Jet Propulsion Laboratory, California Institute of Technology,
under contract with the National Aeronautics and Space Administration.

\label{lastpage}

\end{document}